\documentclass[aps,prl,twocolumn]{revtex4}
\usepackage{graphicx}
\begin{document}

\title{Simulations of composite carbon films with nanotube inclusions}

\author{M. G. Fyta and P. C. Kelires}
\affiliation{Physics Department, University of Crete, P.O. Box 2208, 
710 03, Heraclion, Crete, Greece}

\date{\today}

\begin{abstract}

We study the interfacial structure, stability, and elastic properties 
of composite carbon films containing nanotubes. Our Monte Carlo
simulations show that Van der Waals forces play a vital role in
shaping up the interfacial geometry, producing a curved graphitic wall 
surrounding the tubes. The most stable structures are predicted to have 
intermediate densities, high anisotropies, and increased elastic 
moduli compared to pure amorphous carbon films.
\end{abstract}
\maketitle


Interest on nanocomposite carbon systems is steadily growing for both
practical and fundamental reasons. These materials combine the
properties of carbon nanostructures with those of the 
embedding medium, which usually is an amorphous carbon (a-C) 
matrix. By appropriately choosing the type, size, 
and positioning of the nanostructures, one aims at controlling the 
mechanical and optoelectronic properties of the composite system.

From a fundamental point of view, it is vital to understand how the 
nanostructures interact with the embedding matrix, to unravel the 
structural elements at the interface, and to examine their influence on the 
properties of the material. A very interesting case 
arises when the composite contains nanostructures which, in the absence
of a matrix, are assembled by long-range Van der Waals (VdW) forces.
For example, films containing nanotube fragments and fullerene-like 
inclusions have been produced, and reported to have high hardness and 
high elastic recovery.\cite{Amaratunga} One of the unknown factors in this 
case is how the matrix, a purely covalent material, bonds to the 
nanostructure. This issue, although important for the optimization of 
nanocomposite films, has not been addressed so far at the atomistic level.

We report here the first direct simulations of a prototypical nanocomposite 
material, made up of carbon nanotubes (CNT) inside an a-C matrix. Our
goal is to clearly identify the interfacial structure, and to infer the
stability and hardness of the material. A computational
approach to this problem needs to take into account not only
short-range covalent forces, which are sufficient for the description of 
the interactions within either the CNT's and the a-C matrix, but also
the weaker long-range VdW forces, which might be important for the 
interactions between the CNT's and the matrix. In order to capture the 
subtle effects of such forces on the structure of the system, we need to 
consider simulational cells of realistic dimensions (thousands of atoms.)
This excludes the use of either {\it ab initio} or tight-binding schemes. 

We therefore utilize a less sophisticated computational scheme based on 
empirical potentials, which is however capable of describing large systems 
and correctly introducing VdW forces. For the intra-tubule and 
intra-matrix short-range interactions, we use the Tersoff 
potential,\cite{Terspot} which provides a fairly good description of the 
structure and energetics of a wide range of carbon 
materials.\cite{Ters-Vander,Kel94-00,Fyta03} We confirmed that the
potential reproduces the high elastic moduli of single-wall carbon 
nanotubes (SWCN's). For example, individual SWCN's with diameters of 
$\sim$ 1 nm exhibit bulk moduli of $\sim$ 200 GPa, in very good agreement 
with {\it ab initio} results.\cite{Ordejon} 

For the interactions between the CNT's and the matrix, we use a simple
Lennard-Jones potential,\cite{Martin} which was shown to succesfully 
describe the bulk properties of solid $C_{60}$\cite{Martin} and 
multiple-shell carbon fullerenes.\cite{Lu-Yang} We limit the  
attractive interactions between atoms of the nanotube and atoms in the 
matrix within a cutoff distance of 0.8 nm. Interactions beyond this
distance contribute negligibly to the energy.

\begin{figure}
\includegraphics*[width=7.5cm]{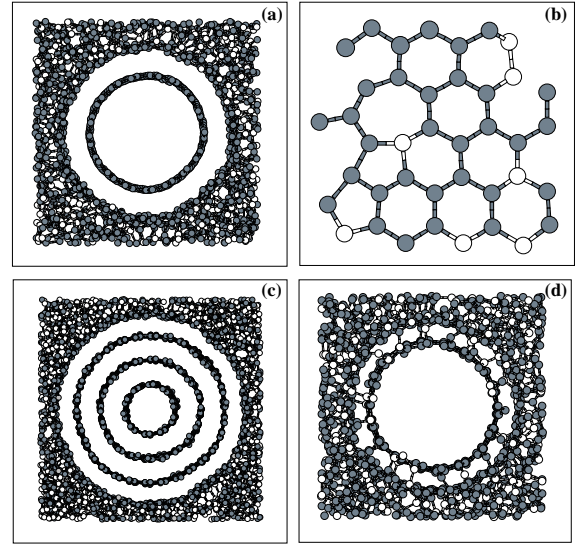}
\caption{Cross sections from ball and stick models of carbon nanotube 
composite structures. (a) An embedded SWCN. (b) Part of the curved graphitic
wall surrounding the SWCN. (c) An embedded MWCN. (d) A structure formed
without VdW interactions. Filled (open) spheres denote sp$^2$ (sp$^3$)
atoms.}
\end{figure}

The composite structures are generated using a continuous-space Monte Carlo 
(MC) approach, extensively described in the past.\cite{Kel94-00,Fyta03}
The initial structure to start with is a cubic diamond crystal with an
empty cylindrical core of a predetermined diameter, formed by artificially 
removing the atoms within this volume, into which the chosen CNT is 
inserted. The amorphous matrix is then generated by melting and subsequent
quenching of the surrounding diamond atoms, while keeping the atoms 
of the CNT frozen in their ideal positions. Finally, full relaxation of the 
whole structure takes place, both in the atomic positions and the 
volume (density) of the cell. Properties are calculated by MC ensemble
averaging at 300 K.

Matrices of various densities (mean coordination) are formed by 
appropriately choosing the size of the initial cell. The CNT's
have open ends and extend through the entire length of the cube.
Because periodic boundary conditions (PBC) are applied to the cells, 
this corresponds to CNT's of infinite length. The size of the cells ranges 
from about 1.5 to 4.0 nm. Nanotubes of various chiralities were embedded, 
with diameters ranging from about 0.8 to 2.8 nm.
Due to the PBC, this corresponds to a dense array of CNT's packed
in parallel. Although this probably is an idealized model of a CNT 
nanocomposite, it provides the essential features of the CNT--matrix 
interaction. The system resembles CNT bundles, but with a-C material 
in between the tubes. 

Representative nanocomposite structures formed this way are portrayed 
in Fig. 1. Panel (a) shows a cross section of a (7 $\times$ 10) SWCN, having
a diameter of 1.2 nm, embedded in a matrix with mean coordination 
$\bar{z}$ = 3.24 (density = 2.29 gcm$^{-3}$). The remarkable feature in
the structure is that the atoms of the 
surrounding matrix reconstruct in such a way as to form an outer wall 
concentric to the nanotube. A double-wall nanotube is
effectively formed. The outer wall is a curved graphitic sheet, but with
some degree of disorder, as a close inspection of its structure reveals. 
Part of this sheet is depicted in panel (b). A ring statistics analysis 
shows that the sheet mainly consists of six-fold rings, their fraction 
being considerably higher than the corresponding fraction in the rest of 
the matrix. However, odd-membered rings do exist, necessitated by the 
presence of some sp$^3$ atoms in the wall, which serve as bridges with 
the rest of the matrix.

In the lowest-energy structures (see below), the optimum distance between 
the outer wall and the CNT is close to the graphite interplanar one 
(0.34 nm). In less stable structures, it deviates from this value depending
on the matrix density. This distance is independent of the tube diameter.
A similar pattern is also observed in structures with embedded 
multi-wall carbon nanotubes (MWCN's), as shown in Fig. 1(c). The interaction 
of the MWCN with the matrix forms again a graphitic sheet. The outer 
interplanar distance is the same with the inner interwall separations in 
the MWCN. The phenomenon is independent of the number of sheets in the MWCN.

Note that in all cases the interaction of the matrix 
graphitic shell with the CNT's does not involve any covalent bonding.
In order to unravel the contributions of short- and long-range forces
to the formation of this wall, we generated structures by only using 
the short-ranged Tersoff potential for the interactions through out the 
system, thus ``turning off'' the VdW interactions. One of
these structures is shown in Fig. 1(d). Again, an outer wall is being
formed, but it is defective. Many covalent bridge bonds are
generated between the matrix wall and the SWCN, and so the interwall 
distance necessarily approaches the value of $\sim$ 0.15 nm, much smaller 
than the interplanar distance in graphite or in MWCN's, and close
to the covalent bond length value.

This indicates that the repulsive forces of the potential tend to drive the 
matrix and CNT atoms apart, but its attractive forces impose these artificial 
bridge bonds, which necessitate the transformation of ideal sp$^2$ sites 
on the nanotube surface into distorted sp$^3$ sites, as shown
in the plot. We conclude that 
short-range attractive forces are not involved in the matrix--CNT 
interaction, and that the VdW forces, although weak,\cite{VdWener} 
are driving the perfect reconstruction of the matrix atoms while keeping
intact the ideal CNT geometry.

\begin{figure}
\begin{center}
\includegraphics*[width=6.3cm]{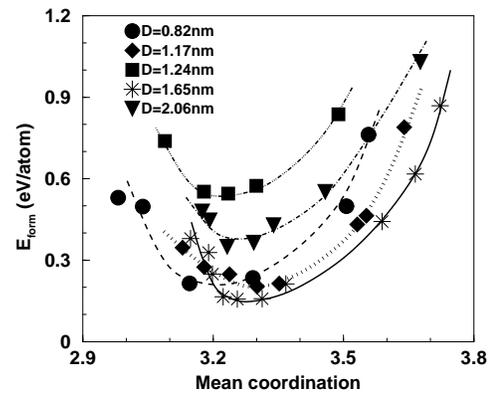}
\end{center}
\caption{Formation energy of SWCN's of various diameters versus mean 
coordination of the embedding matrix. Lines are fits to
the points.}
\end{figure} 

Stability is a crucial property of any composite system. As a measure of
the stability of CNT's in a-C, we calculated their formation 
energy,\cite{Fyta03} defined as the energy of the whole structure compared 
to the sum of the energies of its constituents (the a-C matrix and the 
nanostructure.) The formation energies E$_{form}$ of a number of SWCN's 
of various diameters versus the matrix coordination are plotted 
in Fig. 2. Well defined minima exist for matrices of $\bar{z} \simeq$ 
3.2 -- 3.3, corresponding to densities of about 2.3 -- 2.4 g cm$^{-3}$. 
The denser (or diluted) the matrix the less stable the system becomes. 
Indeed, annealing at elevated temperatures results in heavy distortion 
of CNT's when embedded in dense (diluted) matrices, while nanotubes in 
matrices with $\bar{z}$ near the minima retain their optimal shape.

For dense matrices, this result is understood by noting that the denser 
the a-C matrix the fewer the sp$^2$ sites, and the formation of a graphitic 
wall to interact {\it via} VdW forces with the CNT becomes less easy. In
fact, the wall in such cases is higly defective, with many sp$^3$ sites,
and its distance to the CNT shrinks to $\sim$ 0.24 -- 0.28 nm. The
incomplete wall -- CNT interaction results in CNT distortions. Both
deformations in the wall and the CNT increase the energy.
Diluted a-C matrices, on the other hand, are characterized by a sizeable
fraction of sp$^1$ sites, which act as defects on a graphitic plane.
We find an increasing number of sp$^1$ sites on the wall with decreasing
density. This again destroys locally the curved graphitic nature,
increasing the energy.

\begin{figure}
\begin{center}
\includegraphics*[width=5.2cm,angle=270]{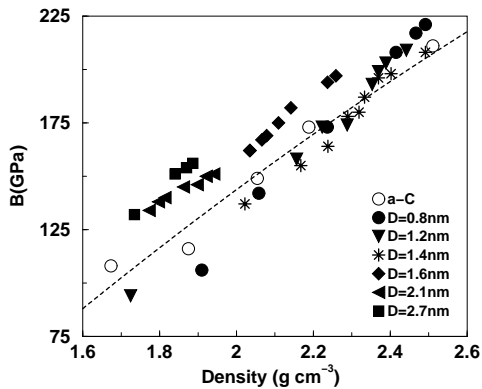}
\end{center}
\caption{Bulk moduli of SWCN's of various diameters versus 
nanocomposite density. Dashed line is a fit to a-C values.}
\end{figure} 

Finally, we have examined the elastic properties of this 
nanocomposite system. A representative quantity is the bulk modulus 
$B$. We calculated $B$ for several structures containing CNT's of various 
diameters. Fig. 3 shows the calculated moduli plotted as a function of the 
nanocomposite density. As a comparison, calculated moduli of single-phase 
a-C films are also shown. We observe a systematic 
increase of the nanocomposite $B$, with respect 
to a-C, when it contains tubes of larger diameter, while smaller tubes 
induce increase only when embedded in dense matrices.

This behavior is the result of a compromise between two
competing factors, i.e., the matrix density and the tube diameter. 
Because we have a high density of tubes in the medium, their size
becomes a crucial factor. The larger the tube diameter the smaller the 
contribution of the matrix to the material's {\bf $B$}, and vice versa. 
By having larger tubes at low densities, the contribution of the soft
matrix declines, and that of the tubes strengthens the material.
Smaller embedded tubes operate in the opposite direction, despite the
fact that the moduli of isolated CNT's increase as their diameter 
decreases. In more dense and rigid matrices, inclusion of smaller tubes 
enhances {\bf $B$} both because such tubes are very rigid and because the
matrix contribution dominates. Also, our calculations show 
that {\bf $B$} is independent of CNT chiralities.

Another marked characteristic of these nanocomposites is 
their high elastic anisotropy. We show this by calculating the 
components of the Young's modulus in the direction of the tube (axial) 
$Y_{axial}$ and in the transverse directions $Y_{trans}$. We find that 
in all cases the anisotropy, defined by the ratio 
$A = Y_{trans} / Y_{axial}$, 
approaches that for the isolated SWCN. For example, a SWCN with a
diameter of 0.8 nm has $Y_{axial}$ = 1100 GPa and $Y_{trans}$ = 620 GPa,
yielding $A$ = 0.56, while the nanocomposite with this tube and a matrix of
density 2.6 gcm$^{-3}$ has $Y_{axial}$ = 570 GPa, $Y_{trans}$ = 380 GPa,
and $A$ = 0.67. The anisotropy increases with increasing tube diameter
(tube volume fraction), because the tube contribution overwhelms the 
isotropic matrix part. Another measure of elastic anisotropy is provided
by the factor $\eta = 2C_{44}/(C_{11}-C_{12})$. A value of unity corresponds 
to a completely isotropic medium. By calculating the elastic constants for
a number of nanocomposites, we confirmed their high anisotropy. For example,
the structure shown in Fig. 1(a) has $\eta$ = 0.75. All pure a-C networks
have $\eta \simeq$ 0.98. A more detailed account of the interesting 
elastic properties of embedded tubes will be given elsewhere.

The interplay between the nanotube size and the embedding density
gives one the opprortunity to tailor the mechanical properties of this
nanocomposite system. It is striking that higher moduli are
achieved by having the tubes in a ``bundle'' arrangement without the
need to interlink them.\cite{Amaratunga} In addition, the CNT composite 
material exhibits much higher bulk moduli than nanotube bundles 
($\sim$ 40 GPa),\cite{Ordejon} and is expected to also have high elastic 
recovery. These superior mechanical properties make it suitable for many 
practical applications.

This work is supported by a grant from the EU and the Ministry of National
Education and Religious Affairs of Greece through the action
``E$\Pi$EAEK'' (programme ``$\Pi\Upsilon\Theta$A$\Gamma$OPA$\Sigma$''.)


\begin{references}

\bibitem{Amaratunga} G. A. J. Amaratunga, M. Chhowalla, C. J. Kiely,
I. Alexandrou, R. Aharonov, and R. M. Devenish, Nature {\bf 383}, 321 (1996).
\bibitem{Terspot} J. Tersoff, Phys. Rev. Lett. {\bf 61}, 2879 (1988). 
\bibitem{Ters-Vander} D. Vanderbilt and J. Tersoff, Phys. Rev. Lett. 
{\bf 68}, 511 (1992).
\bibitem{Kel94-00} P. C. Kelires, Phys. Rev. Lett. {\bf 73}, 2460 (1994);
Phys. Rev. B {\bf 62}, 15686 (2000).
\bibitem{Fyta03} M. G. Fyta, I. N. Remediakis and P. C. Kelires,  
Phys. Rev. B {\bf 67}, 035423 (2003).
\bibitem{Ordejon} S. Reich, C. Thomsen, and P. Ordej\'{o}n, 
Phys. Rev. B {\bf 65}, 153407 (2002).
\bibitem{Martin} J. P. Lu, X. -P. Li and R. M. Martin, Phys. Rev. Lett. 
{\bf 68}, 1551 (1992).
\bibitem{Lu-Yang} J. P. Lu and W. Yang, Phys. Rev. B {\bf 49}, 11 421 (1994).
\bibitem{VdWener} We estimate the contribution of VdW forces to the 
cohesive energy of the system to be $\sim$ 0.2\%.

\end{references}
\end{document}